\documentclass[prl,aps,twocolumn,showpacs]{revtex4}
\usepackage{graphics}
\begin{document}

\title{
Exciton-Population Inversion and Terahertz Gain 
in Resonantly Excited Semiconductors}

\author{M. Kira}
\author{S. W. Koch}
\affiliation{Fachbereich Physik and Material Sciences Center, 
Philipps Universit\"at, Renthof 5, 35032 Marburg/Germany}

\date{\today}

\begin{abstract}
The build-up of exciton populations in resonantly laser
excited semiconductors is studied microscopically. For excitation at the
$2s$-exciton resonance, it is shown that
polarization with a strict $s$-type radial symmetry
can be efficiently converted
into an incoherent $p$-type population. As a consequence, inversion
between the $2p$ and $1s$ exciton states can be obtained leading to
the appearance of significant terahertz gain. 
\end{abstract}


\maketitle

\noindent

Resonant laser excitation of semiconductors
induces a coherent interband polarization between
conduction-band electron and valence-band hole states.
Through interaction and scattering processes this optical
polarization may be converted into incoherent populations
of unbound or bound electron-hole pairs (excitons).
Even though excitonic features of the coherent polarization
are well understood \cite{elliott,Haug:04}, the study of its decay into
incoherent many-body states is an area of
active research. 

Several recent experiments have applied terahertz (THz) fields
to directly probe the optically
generated many-body system.
So far, this approach has been used to 
detect and monitor 
conductivity \cite{Beard:00},
plasmons \cite{Huber:01},
and bound exciton formation \cite{excitons}. At the same time,
tunable terahertz (THz) sources evolve rapidly ranging from
free-electron \cite{free-electron} and  
quantum-cascade lasers \cite{QCL} to sources with
difference-frequency generation \cite{Kleine:01}.
Since the semiconductor quasi-particle excitations
strongly interact with THz radiation, it is an interesting question
to see if an excitonic system could actually be used to generate
THz radiation, or even to provide THz amplification, i.e. THz gain.

In order to explore the THz properties of resonantly
excited semiconductors, we study the build-up of
exciton populations in different quantum states.  
Besides the conversion of excitonic polarizations
at the $1s$ or $2s$ resonances into incoherent 
$s$-type populations, we show that {\it Coulomb induced
scattering can efficiently convert excitonic coherences
with a strict $s$-type radial symmetry
into an incoherent $p$-type population}. 
We show that this process may even lead to a 
population inversion between the $2p$ and $1s$ states, 
giving rise to THz probe gain.

As a model system, we analyze
quantum-wire structures but we show that the
main results are equally valid for quantum-well systems.
The electronic excitations are
described by Fermion operators 
$a_{c(v),{\bf k}}$ and $a^\dagger_{c(v),{\bf k}}$
related to an electron with carrier momentum ${\bf k}$ 
in the conduction (valence) band. We include the 
carrier-carrier Coulomb interaction as well as 
the couplings to light fields and phonons \cite{Haug:04}.
The dynamic evolution of single-particle quantities related to the
microscopic polarization
$P_{\bf k} \equiv \langle a^\dagger_{v,{\bf k}} a_{c,{\bf k}} \rangle$,
electron 
$f^{e}_{\bf k} \equiv \langle a^\dagger_{c,{\bf k}} a_{c,{\bf k}} \rangle$,
and hole 
$f^{h}_{\bf k} \equiv \langle a_{v,{\bf k}} a^\dagger_{v,{\bf k}}\rangle$
densities obey the semiconductor Bloch equations \cite{Haug:04} 
\begin{eqnarray}
  &&i \hbar \frac{\partial}{\partial t} 
  P_{{\bf k}} 
  = 
  [\epsilon_{{\bf k}} + j_{\bf k} A]
  P_{{\bf k}} 
- \left[1-f^e_{{\bf k}}-f^h_{{\bf k}}\right] \Omega_{{\bf k}} 
  -i\Gamma_{\bf k}  \quad  
\label{eq:SBE-Pol}
\\
   &&\frac{\hbar}{2}\frac{\partial}{\partial t}
   f^e_{\bf k}
   =
     {\rm Im}
	\left[
   P_{\bf k}
   \Omega^{\star}_{\bf k}
   +\sum_{{\bf q},{\bf k}',\lambda} V_{{\bf q}}
     c^{{\bf q},{\bf k}',{\bf k}}_{c,\lambda,\lambda,c}
   +\sum_{\bf q} 
 	{\cal D}^{\rm c,c}_{{\bf k},{\bf q}}
   \right]  
\label{eq:fe_dyn}
\\
  &&i\Gamma_{\bf k} =
  \sum_{\bf q}
	\left[
	V_{{\bf q}}
	\sum_{{\bf n},\lambda}
	c^{{\bf q},{\bf n},{\bf k}}_{v,\lambda,\lambda,c}
	-
 	{\cal D}^{\rm v,c}_{{\bf k},{\bf q}}
	\right]
     -
     \left[
	c \leftrightarrow v
	\right]^\star
\label{eq:P_decay}
\end{eqnarray}
and similarly for $f^h_{\bf k}$. Here, $V_{\bf k}$ is
the Coulomb-matrix element whereas $\epsilon_{{\bf k}}$
and $\Omega_{\bf k}$ denote the renormalized kinetic
energy and Rabi frequency, respectively. 
The coupling to the THz field follows from the
$j A$ term which contains the vector potential 
$A {\bf e}_\sigma$ with direction
${\bf e}_{\sigma}$ and
 the current-matrix element 
$j_{\bf k} = e(\frac{1}{m_e} + \frac{1}{m_h}) 
\hbar {\bf k} \cdot {\bf e}_{\sigma}$
with the electron (hole) mass $m_{e(h)}$.
The true two-particle correlations stem from
$c_{\lambda,\nu,\nu',\lambda'}^{{\bf q},{\bf k}',{\bf k}} \equiv
\Delta 
\langle a^{\dagger}_{\lambda,{\bf k}} a^{\dagger}_{\nu,{\bf k}'}  
a_{\nu',{\bf k}'+{\bf q},}  a_{\lambda',{\bf k}-{\bf q}} \rangle$, 
and
${\cal D}^{\lambda,\nu}_{{\bf k},{\bf q}} =
\sum_{p_z} {\cal G}_{p_z,{\bf q}} \Delta \langle
\left( D_{p_z,{\bf q}} + D^{\dagger}_{p_z,{\bf q}} \right)
a^{\dagger}_{\lambda,{\bf k}} a_{\nu,{\bf k} - {\bf q}} \rangle$
where $D$ and $D^\dagger$ are the bosonic phonon operators
and ${\cal G}_{p_z,{\bf q}}$ is the phonon-matrix element.
The notation $\Delta$ indicates that 
the factorized single-particle contributions 
(subscript $S$) are removed,
e.g. $\Delta 
\langle a^{\dagger}_{\lambda,{\bf k}} a^{\dagger}_{\nu,{\bf k}'}  
a_{\nu',{\bf k}'+{\bf q},}  a_{\lambda',{\bf k}-{\bf q}} \rangle
=
\langle a^{\dagger}_{\lambda,{\bf k}} a^{\dagger}_{\nu,{\bf k}'}  
a_{\nu',{\bf k}'+{\bf q},}  a_{\lambda',{\bf k}-{\bf q}} \rangle
- 
\langle a^{\dagger}_{\lambda,{\bf k}} a^{\dagger}_{\nu,{\bf k}'}  
a_{\nu',{\bf k}'+{\bf q},}  a_{\lambda',{\bf k}-{\bf q}} 
\rangle_{\rm S}$.
The two-particle correlations not only influence the population dynamics, they
also determine $\Gamma_{\bf k}$ which leads to the decay of $P$.

If one ignores $\Gamma_{\bf k}$, the optically generated state 
$| \Psi_{\rm coh} (t) \rangle$ can easily be obtained
from the well-known coherent-limit results \cite{Haug:04}  
of Eqs.~(\ref{eq:SBE-Pol}) - (\ref{eq:fe_dyn}).  
For excitations resonant with the excitonic state 
$\phi_\nu({\bf k})$, we find
$P_{\bf k} \propto \phi_\nu({\bf k}) \sum_{{\bf k}'} \phi_\nu({\bf k}')$
and the conservation law
$(n_{\bf k} - \frac{1}{2})^2 + |P_{\bf k}|^2 = \frac{1}{4}$
where $n_{\bf k} \equiv f^{e}_{\bf k} = f^{h}_{\bf k}$.
These {\it coherent excitons} have 
a strict $s$-like symmetry since only then 
$\sum_{{\bf k}'} \phi_\nu({\bf k}')$ is nonvanishing.
In addition, $| \Psi_{\rm coh} (t) \rangle$ is a Slater determinant
$  | \Psi_{\rm coh} (t) \rangle
  = \prod_{\bf k} L^{\dagger}_{\bf k}(t) | \Psi_{0} \rangle
  = D(\hat{X})
    \prod_{\bf k} a^{\dagger}_{v,{\bf k}} | \Psi_{0} \rangle $
where $| \Psi_{0} \rangle$ is the unexcited semiconductor
while
$L^{\dagger}_{\bf k}(t) = 
{\rm e}^{i\varphi_{\bf k}(t)} {\rm sin} \beta_{\bf k}(t) 
\; a^{\dagger}_{c,{\bf k}}
+{\rm cos} \beta_{\bf k}(t) \; a^{\dagger}_{v,{\bf k}}$
is a fermion operator with
${\rm sin}^2 \beta_{\bf k}(t) = n_{\bf k}(t)$
and
${\rm e}^{i\varphi_{\bf k}(t)} = P_{\bf k}(t) / |P_{\bf k}(t)|$.

For our subsequent discussions, it is convenient to introduce an exciton
operator $
  \hat{X}_{\nu,{\bf q}} = \sum_{\bf k} \phi_{\nu}({\bf k}) 
	a^{\dagger}_{v,{\bf k}-{\bf q}_h} 
	a_{c,{\bf k}+{\bf q}_e} $
with a center-of-mass momentum
${\bf q}$ and ${\bf q}_{e(h)} = m_{e(h)} /(m_e+m_h){\bf q}$.
By choosing $\hat{X} = \hat{X}_{\nu,{\bf 0}}$
and 
$\phi_\nu({\bf k}) = \beta_{\bf k} e^{i\varphi_{\bf k}}$,
$| \Psi_{\rm coh} (t) \rangle$
follows from
$D(\hat{X}) = e^{\hat{X}^{\dagger}-\hat{X}}$
acting on the full valence band.
Since $D(\hat{X})$ is formally analogous to the coherent 
state generator of bosonic fields \cite{Walls:94}, 
we may interpret $| \Psi_{\rm coh} (t) \rangle$
as a coherent exciton state
even though $\hat{X}$ is not bosonic \cite{Usui:60}.

To study, how efficiently $| \Psi_{\rm coh} (t) \rangle$ 
can be converted into incoherent excitons, we need to solve the full 
Eqs.~(\ref{eq:SBE-Pol})-(\ref{eq:P_decay}).
For resonant excitation, both  $c_{v,\lambda,\lambda,c}$
and ${\cal D}_{{\bf k},{\bf q}}$
convert a coherent excitonic
polarization into incoherent two-particle populations 
where
$c^{{\bf q},{\bf k}',{\bf k}}_{\rm X} \equiv 
c^{{\bf q},{\bf k}',{\bf k}}_{c,v,c,v}$
describes incoherent excitons. 
Using the cluster expansion \cite{cluster},
we derive equations
for the two-particle correlations:
\begin{eqnarray}
  &&i \hbar \frac{\partial}{\partial t}
  c^{{\bf q},{\bf k}',{\bf k}}_{\rm X}
= (\epsilon^{{\bf q},{\bf k}',{\bf k}} 
   +j_{{\bf k}'+{\bf q}-{\bf k}}A)
c^{{\bf q},{\bf k}',{\bf k}}_{\rm X}
+ S^{{\bf q},{\bf k}',{\bf k}} \quad \quad
\nonumber\\
  &&\;\;\;\;+
	(1-f^e_{{\bf k}}-f^h_{{\bf k}-{\bf q}})
	\sum_{\bf l} V_{{\bf l}-{\bf k}}
   		c^{{\bf q},{\bf k}',{\bf l}}_{\rm X}
\nonumber\\
  &&\;\;\;\;-
  (1-f^e_{{\bf k}'+{\bf q}}-f^h_{{\bf k}'})
	\sum_{\bf l} V_{{\bf l}-{\bf k}'}
      	c^{{\bf q},{\bf l},{\bf k}}_{\rm X}
\nonumber\\
&&\;\;\;\;
+iG^{{\bf q},{\bf k}',{\bf k}}
+D^{{\bf q},{\bf k}',{\bf k}}_{\rm rest}
+T^{{\bf q},{\bf k}',{\bf k}}
\label{eq:EXPcvcv},
\\
&&iG^{{\bf q},{\bf k}',{\bf k}} =
   	(P^{\star}_{\bf k} - P^{\star}_{{\bf k}-{\bf q}}) 
	V_{\bf q}
	\left[
   	\sum_{{\bf n},\lambda}
   		c^{-{\bf q},{\bf n},{\bf k}'}_{v,\lambda,\lambda,c}
		-
		{\cal D}^{v,c}_{{\bf k}',-{\bf q}}
	\right]
\nonumber\\
  &&\;\;\;\;\;\;\;\;\;\;\;\;\;\;\;\;
  +(P_{{\bf k}'} - P_{{\bf k}'+{\bf q}}) 
	V_{\bf q}
	\left[
	\sum_{{\bf n},\lambda}
	c^{{\bf q},{\bf n},{\bf k}}_{c,\lambda,\lambda,v}
	-
	{\cal D}^{c,v}_{{\bf k},{\bf q}} 
    \right]
\label{eq:Xgeneration},
\end{eqnarray}
where $\epsilon^{{\bf q},{\bf k}',{\bf k}}$ is
the renormalized kinetic energy of the
two-particle state and $S$ contains
Coulomb induced in-and-out scattering of single-particle
quantities. The Coulomb sums with
the phase-space filling factor
$(1-f^e-f^h)$ describe
the attractive interaction between electrons and holes,
allowing them to become truly bound electron-hole pairs,
i.e.~{\it incoherent excitons} which can be probed via the
THz induced  $j\cdot A$ coupling. The $G$ term contains
the same  $c_{v,\lambda,\lambda,c}$ and ${\cal D}^{v,c}$ 
correlations as $\Gamma$ in Eq.~(\ref{eq:P_decay}),
showing how coherent excitons are converted into incoherent ones. 
The remaining two-particle contributions are denoted as $D_{\rm rest}$
while $T$ 
symbolizes the three-particle Coulomb and phonon terms treated here
at the scattering level.
This way, we fully include
one- and two-particle correlations and obtain a closed set
of equations providing a consistent description
of optical as well as THz excitations in semiconductors.

The polarization to population conversion efficiency is determined
from the density of incoherent $\nu$-excitons \cite{Kira:01}
$\Delta n_{\nu} = \frac{1}{{\cal L}^{d}} \sum_{\bf q} 
		\Delta \langle \hat{X}^{\dagger}_{\nu,{\bf q}} 
		\hat{X}_{\nu,{\bf q}} \rangle
        =
	\sum_{{\bf k},{\bf k}'} 
	\phi^{\star}_\nu({\bf k}) \phi_\nu({\bf k})
	c^{{\bf q},{\bf k}'-{\bf q}_h,{\bf k}+{\bf q}_e}_X
$
where ${\cal L}^{d}$ is the normalization volume.
We evaluate the conversion efficiency by numerically integrating the
complete set of equation for a planar arrangement of identical
quantum wires. 
Later on, we estimate the conversion efficiency also for
a quantum well.
We choose standard GaAs-type parameters
and the wire and well sizes are taken such that the energy separation between 
the two lowest exciton states is 5~meV. The lattice temperature is assumed 
to be 10~K such that it is sufficient to include only acoustic 
phonons\cite{Thraen:00}.
To study the generation of incoherent
excitons in their different quantum states, we
assume 4~ps pulsed optical excitation resonant with either the
$1s$- or $2s$-resonance. We repeat the computations for different pump
intensities and evaluate the final quasi-stationary
exciton fraction $\Delta n_{\nu} / n$
relative to the generated carrier density 
$n = \frac{1}{{\cal L}^d} \sum_{k} f^{e(h)}_{\bf k}$.

\begin{figure}[h]
\center{\scalebox{0.4}{\includegraphics{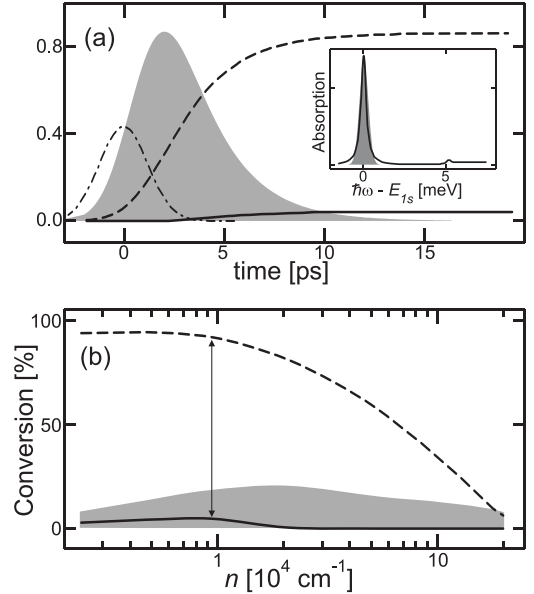}}}
\caption{
(a) For excitation at the $1s$ exciton resonance 
with a $4~ps$ laser pulse (dot-dashed line), the temporal  
evolution of the induced optical polarization $|P|^2$ (shaded area),
together with the generated incoherent $1s$ (dashed line) and 
$2p$ (solid line) exciton densities [$10^4~{\rm cm}^{-1}$] are
shown. The inset shows the pump (shaded area) and
linear absorption (solid line) spectra;  $E_{1s}$ is the 
$1s$-exciton energy. 
(b) The polarization to population
conversion efficiency for $1s$ (dashed line) and $2p$ excitons
(solid line) is plotted as function of excitation density $n$.
The arrow indicates the density at which the dynamics is shown in a).
The shaded area represents the result obtained without the
phonon scattering.
} \label{fig1}
\end{figure}

\begin{figure}[h]
\center{\scalebox{0.4}{\includegraphics{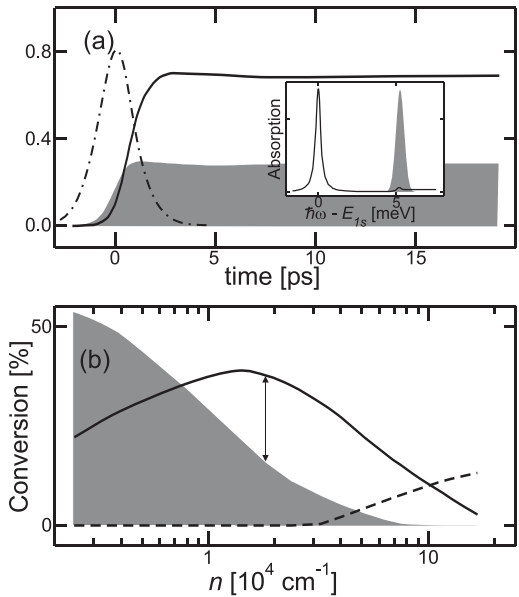}}}
\caption{Same as Fig~.1 but for excitation at the $2s$
exciton resonance (inset). 
(a) Dynamics of optical polarization $|P|^2$ (dot-dashed line)  
and incoherent densities of $2s$ (shaded area) and 
$2p$ (solid line) excitons [$10^4~{\rm cm}^{-1}$]. 
(b) Conversion efficiency for 
$1s$ (dashed line), $2p$ (solid line),
and $2s$ (shaded area) excitons as function of excitation density $n$.
} \label{fig2}
\end{figure}

In Figs.~\ref{fig1} and \ref{fig2}, we present numerical results for
pumping at the $1s$- and $2s$-resonances, respectively. 
The insets show the spectral excitation conditions.
In Fig.~\ref{fig1}(a),
we plot the temporal evolution of the pump pulse, 
of the induced optical polarization, and 
of the generated 1$s$ and 2$p$ exciton density.
Figure \ref{fig1}(b) presents
the relative percentage of excitons in the different quantum states showing,
not surprisingly, 
that for $1s$ excitation the optical polarization is mainly 
converted into incoherent $1s$ excitons;
$\Delta n_{1s} / n$ is well above 90~\% for
low densities. This large conversion fraction is expected since
coherent and incoherent $1s$ excitons 
have an excellent energetic match. However,
the generated exciton population drops well below 40~\% already
at moderate densities above  $10^5$ ${\rm cm}^{-1}$ 
where the phase space filling factor  $(1 - f^e - f^h)$ peaks around  0.5.
For even higher densities, $\Delta n_{\nu} / n_{eh}$ vanishes
since excitons start to ionize.
A computation, where phonon scattering
${\cal D}$ is omitted, indicates that Coulomb scattering alone would lead 
only to roughly 15~\% exciton population.
 
The $2s$-excitation results are presented in Fig.~\ref{fig2} where
Fig.~\ref{fig2}(b) shows $\Delta n_{2s} / n$,
$\Delta n_{2p} / n$, and $\Delta n_{1s} / n$. For not too high
excitation densities, we
observe that {\it the $2s$-polarization is converted into
a mix of $2s$ and $2p$ populations}. 
Whereas the amount of $2s$ population decreases monotonously
with increasing excitation, the $2p$ population
first increases up to 40~\% before it also decreases at 
higher densities where formation of $1s$-excitons
gradually becomes relevant. 

\begin{figure}[h]
\center{\scalebox{0.45}{\includegraphics{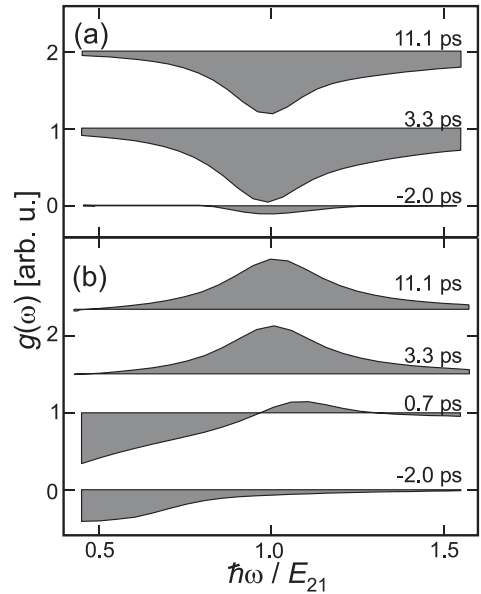}}}
\caption{
Terahertz gain spectra $g(\omega)$ for different time delays 
corresponding to the conditions of (a) 
$1s$ excitation as in Fig.~\ref{fig1}a and (b) 2$s$ excitation as in
Fig.~\ref{fig2}a. All curves are identically scaled but  
shifted with respect to another. Here,
$E_{21}$ is the energy difference between the $1s$ and $2p$ exciton
states.
} \label{fig3}
\end{figure}

Before we analyze the physical mechanism responsible for the
significant formation of a $2p$-exciton population, we study
its signatures assuming a THz probe. For this purpose we
evaluate the $j \cdot A$ terms in Eqs.~(\ref{eq:SBE-Pol}) 
and (\ref{eq:EXPcvcv}) to compute the generated THz current
$J_{\rm THz} = \frac{1}{{\cal L}^d} 
\sum_{{\bf k},\lambda} j_\lambda({\bf k}) f^{\lambda}_{\bf k}$
with both coherent and incoherent contributions \cite{Kira:04}.
We determine the linear THz gain
$g(\omega) = -{\rm Im} \left[ J_{\rm THz}(\omega) 
/ \left( \omega A(\omega) \right) \right]$ 
assuming a 150~fs THz probe pulse capable of resolving 
temporal snapshots of $g(\omega)$ during the exciton formation process.
Figure \ref{fig3}a shows that for $1s$-pumping 
the corresponding $g(\omega)$ is always
negative, i.e. we find THz absorption peaked around 
the $1s$-$2p$ transition. These results, as well as the
asymmetric line shape due to transitions from the $1s$
to energetically higher bound and unbound states
agree well with
experimental findings reported in Ref. \cite{excitons}. 

The same analysis is repeated for the $2s$ excitation
(Fig.~\ref{fig3}b)
showing that $g(\omega)$ rapidly
changes from absorption to gain.
At very early times, the system consists mainly 
of coherent $2s$ excitons such that only absorptive
transitions from the $2s$ to higher states are present.
However, as incoherent $2p$ excitons are generated,
pronounced THz gain develops at the $2p$ to $1s$ transition
as a consequence of the
population inversion between these states. 

Now, after we have presented examples of the fully numerical
evaluation of our many-body theory, we want to analyze the
relevant physical mechanisms. 
At first sight the significant generation of $p$-type excitons
might be unexpected since it involves a
symmetry change of the optically generated
$s$-type polarization. 
To identify the microscopic origin of this process, we 
numerically study the relative
importance of different contributions to the full theory. 
We find that for $2s$ pumping a switch off of phonon-induced scattering 
leaves the generated exciton fraction practically unchanged. 
Hence, in contrast to $1s$ pumping, the {\it conversion
of a $2s$ polarization predominantly results from Coulomb scattering}.

In order to understand how the Coulomb interaction
induces symmetry changes in the polarization conversion we now 
investigate the scattering ($\Gamma$) and 
conversion ($G$) mechanisms which stem from
the same fermionic correlation
$c_{v,\lambda,\lambda,c} 
= \Delta \langle a^\dagger_v a^\dagger_\lambda a_\lambda a_c \rangle$
between polarization and fermionic density. 

(1) In a first step,
we make the lowest level approximation by replacing the
microscopic $\Gamma_{\bf k}$ by a 
phenomenological decay: $\Gamma_{\bf k} = - \Gamma P_{\bf k}$.
In this case, $G^{{\bf q},{\bf k}',{\bf k}} 
= + 2 P^\star_{\bf k} \Gamma P_{{\bf k}'} \; \delta_{{\bf q},0}$ 
and 
$[P^\star_{\bf k} P_{{\bf k}'}
+\sum_{\bf q} c^{{\bf q},{\bf k}',{\bf k}}_X ]$
is a constant of motion with respect to the scattering.
Therefore, this simple dephasing model
only converts $s$-type polarization  
to $s$-like exciton populations, in contrast to our
microscopic results.

(2) Looking at the process of excitation induced
depahsing \cite{Jahnke:97} we see  
that Coulomb induced dephasing is actually a diffusive 
redistribution of the microscopic polarizations since
$\sum_{\bf k} \Gamma_{\bf k} =0$
and
$\sum_{{\bf k},{\bf k}',{\bf q}} G^{{\bf q},{\bf k}',{\bf k}} =0$.
At the same time, Eqs.~(\ref{eq:P_decay}) and (\ref{eq:Xgeneration}) 
impose a strict microscopic connection between $\Gamma$ and $G$.
In order to analyze the consequences of these fundamental restrictions,
we use a somewhat reduced model by utilizing a simplified structural form
of the second-Born solution of $c_{v,\lambda,\lambda,c}$ \cite{Jahnke:97}
via
$\sum_\lambda c_{v,\lambda,\lambda,c}^{{\bf q},{\bf k}',{\bf k}}
= 
-(\sum_\lambda c_{v,\lambda,\lambda,c}^{{\bf q},{\bf k}',{\bf k}})^\star
=i F(f,P)_{{\bf k}',{\bf q}}(P_{{\bf k}-{\bf q}} - P_{\bf q})$
where $F(f,P)$ is a real-valued, nonlinear
functional of $f$ and $P$
containing the energy and momentum conservation aspects of the
Coulomb scattering. The reduced model implies $
	\Gamma_{\bf k}^{\rm red}
	=-
	\sum_{{\bf q}} U_{\bf q}
	(P_{\bf k} - P_{{\bf k}-{\bf q}})$ and $
	G^{{\bf q},{\bf k}',{\bf k}}_{\rm red}
	=
	(P^\star_{\bf k} - P^{\star}_{{\bf k}-{\bf q}})
	U_{\bf q}
	(P_{{\bf k}'+{\bf q}} - P_{{\bf k}'})$ ,
where $U_{\bf q} = 2 V_{\bf q} 
\sum_{{\bf k}'} F(f,P)_{{\bf k}',{\bf q}}$.
As a result, $\Gamma_{\bf k}^{\rm red}$ removes polarization
from the state $P_{\bf k}$ and redistributes it to 
$P_{{\bf k}-{\bf q}}$ while the rate of conversion to the
exciton state $\nu$ becomes
$G^{{\bf q},\nu,\nu}_{\rm red} = |M_{\nu,{\bf q}}|^2 U_{\bf q}$. Here,
the scattering matrix element is
$M_{\nu,{\bf q}} = \sum_{\bf k} \phi^{\star}_\nu({\bf k})
\left[
P_{{\bf k}+{\bf q}_e} - P_{{\bf k}-{\bf q}_h}
\right]$ indicating that Coulomb scattering leads to the
generation of excitons with
finite momenta whereas no population in the
${\bf q} =0$ state is produced. For 
low to moderate $2s$ excitation,
we may use the approximation $P_{\bf k} \propto \phi_{2s}({\bf k})$.
With the help of the symmetries 
$\phi_{2s}(-{\bf k}) = \phi_{2s}({\bf k})$ 
and 
$\phi_{2p}(-{\bf k}) = -\phi_{2p}({\bf k})$,
we find
$M_{2p,{\bf q}} \propto \sum_{{\bf k},\lambda}
\phi^{\star}_{2p}({\bf k})
\phi_{2s}({\bf k}+{\bf q}_\lambda)$
which is clearly nonzero for ${\bf q} \neq 0$.

For $2s$ pumping, the energy conservation aspects of $U_{\bf q}$ are 
practically the same for $2s$ and $2p$ since these
states are nearly degenerate.
As a result, the overlap of the wavefunctions with shifted arguments
in $M_{\nu,{\bf q}}$ determines the conversion rate
such that $|M_{\nu,{\bf q}}|^2$ 
can be used to estimate the ratio of 
generated $2s$ and $2p$ populations.
Using the low-density exciton wavefunctions,
we obtain a ratio of 1.36 of $2s$ over $2p$ population
for the quantum wire, which is close to the numerical result in
Fig.~\ref{fig2}b. Repeating the same calculations for the quantum
well, we get a ratio 0.99 showing 
that the generation of $p$-like states is strong and qualitatively
similar for quantum wells and wires.

Since the Coulomb interaction conserves the angular momentum, one may
ask how this conservation law is fulfilled when a $2s$ polarization is 
converted into $2p$ excitons. Without THz fields, 
Eq.~(\ref{eq:EXPcvcv}) implies correlations with a functional
dependence
$c^{{\bf q},{\bf k}'-{\bf q}_h,{\bf k}+{\bf q}_e} =
c({|{\bf q}|,|{\bf k}'|,|{\bf k}|},{\rm cos} \varphi_{{\bf k},{\bf q}},
{\rm cos} \varphi_{{\bf k}',{\bf q}})$
where $\varphi_{{\bf k}({\bf k}'),{\bf q}}$ is the angle between
${\bf k}$ (${\bf k}'$) and ${\bf q}$.
Consequently, the generated excitons have an 
angular dependency
${\rm cos} ^m \varphi_{{\bf k},{\bf q}} = 
(e^{+i\varphi_{{\bf k},{\bf q}}}
+e^{-i\varphi_{{\bf k},{\bf q}}})^m/2^m$
such that the eigenfunctions $e^{+im\varphi}$ and $e^{-im\varphi}$
of the $L_z$ state are generated with equal probability.
As a result, the total $\langle L_i \rangle$ always vanishes 
such that the total angular momentum is fully conserved even 
when $2p$ excitons are generated.

In summary, our microscopic study predicts significant formation of
excitons in $2p$ states for excitation at the $2s$ resonance of
the absorption spectrum. As a consequence, an exciton population
inversion between the $2p$ and $1s$ states is realized leading
to gain for the corresponding THz frequency. Using different semiconductor
materials this scheme may become useful for THz amplification in
a wide spectral range.  

\begin{acknowledgments}
This work was supported by the Optodynamics Center and
the Deutsche Forschungsgemeinschaft through
the Quantum Optics in Semiconductors Research Group.
\end{acknowledgments}


\begin{references}

\bibitem{elliott}
R.J. Elliott, in {\it Polarons and Excitons}, eds. C.G. Kuper and 
G.D. Whitefield, Oliver and Boyd, (1963), {\it p}. 269.

\bibitem{Haug:04}
H.~Haug and S.W.~Koch,
{\it Quantum Theory of the Optical and Electronic Properties of Semiconductors}
(World Scientific Publ., Singapore, 4th ed., 2004).
%

\bibitem{Beard:00}
M.C.~Beard {\it et al.},
Phys.~Rev.~B {\bf 62}, 15764 (2000)

\bibitem{Huber:01}
R.~Huber {\it et al.},
Nature~{\textbf{414}}, 286 (2001).

\bibitem{excitons}
J. Cerne {\it et al.},
Phys.~Rev.~Lett.~{\bf 77}, 1131 (1996);
R.A.~Kaindl {\it et al.},
Nature {\bf 423}, 734 (2003).

\bibitem{free-electron}
J.~Urata {\it et al.}, Phys.~Rev.~Lett.~{\bf 80}, 516 (1998);
M.~Abo-Bakr {\it et al.}, Phys.~Rev.~Lett.~{\bf 88}, 254801 (2002).

\bibitem{QCL}
J.~Faist {\it et al.}, Science {\bf 264}, 553 (1994);
B.S.~Williams BS {\it et al.}, Appl.~Phys.~Lett.~{\bf 83}, 5142 (2003).

\bibitem{Kleine:01}
T.~Kleine-Ostmann {\it et al.}, Electronics Lett.~{\bf 37}, 1461 (2001).

\bibitem{Walls:94}
D.F.~Walls and G.J.~Milburn,
{\it Quantum Optics}
(Springer-Verlag, Berlin, 1994).

\bibitem{Usui:60}
T.~Usui, Prog.~Theor.~Phys.~{\bf 23}, 787 (1960).

\bibitem{cluster}
H.W.~Wyld and B.D.~Fried, Ann.~Phys.~{\bf 23}, 374 (1963);
J.~Fricke, Ann.~Phys.~{\bf 252}, 479 (1996);
M.~Kira {\it et al.}, J.~Nonlin.~Opt.~B~{\bf 29}, 481 (2002).

\bibitem{Kira:01}
M.~Kira {\it et al.},
Phys.~Rev.~Lett.~\textbf{87}, 176401 (2001);
W. Hoyer {\it et al.},
Phys.~Rev.~B~{\bf 67}, 155113 (2003)

\bibitem{Thraen:00}
A.~Thr{\"a}nhardt {\it et al.},
Phys.~Rev.~B \textbf{62}, 2706 (2000).

\bibitem{Kira:04}
M.~Kira {\it et al.}, Solid State Commun.~{\bf 129}, 733 (2004).

\bibitem{Jahnke:97}
F.~Jahnke {\it et al.},
Z.~Physik B \textbf{104}, 559 (1997).
%
\end{references}
\end{document}